\documentclass[jgrga]{AGUTeX}

\usepackage{graphicx}
\authorrunninghead{SHEPHERD AND CASSAK}

\titlerunninghead{X-LINE SPREADING}

\authoraddr{P.~A.~Cassak, Department of Physics, White Hall, Box 6315,
  West Virginia University, Morgantown, WV 26506.
  (Paul.Cassak@mail.wvu.edu)}

\authoraddr{L.~S.~Shepherd, Department of Physics, White Hall, Box
  6315, West Virginia University, Morgantown, WV 26506,
  (lshephe1@mix.wvu.edu)}

\begin{document}
\hskip 8.5cm {\it Preprint - Submitted April 2012, Resubmitted July 2012} \vskip .125cm

\title{Guide Field Dependence of 3D X-line Spreading During
  Collisionless Magnetic Reconnection}

\authors{L.~S.~Shepherd\altaffilmark{1} and P.~A.~Cassak\altaffilmark{1}}

\altaffiltext{1}{Department of Physics,
West Virginia University, Morgantown, West Virginia, USA.}

\begin{abstract}
  Theoretical arguments and large-scale two-fluid simulations are used
  to study the spreading of reconnection X-lines localized in the
  direction of the current as a function of the strength of the
  out-of-plane (guide) magnetic field.  It is found that the mechanism
  causing the spreading is different for weak and strong guide fields.
  In the weak guide field limit, spreading is due to the motion of the
  current carriers, as has been previously established.  However,
  spreading for strong guide fields is bi-directional and is due to
  the excitation of Alfv\'en waves along the guide field.  In general,
  we suggest that the X-line spreads bi-directionally with a speed
  governed by the faster of the two mechanisms for each direction.  A
  prediction on the strength of the guide field at which the spreading
  mechanism changes is formulated and verified with three-dimensional
  simulations.  Solar, magnetospheric, and laboratory applications are
  discussed.
\end{abstract}

\begin{article}

\section{Introduction}
Magnetic reconnection is the basic plasma process in which magnetic
energy is converted to kinetic and thermal energies (\cite{Dungey53,
  Vasyliunas75}).  It plays an important role in the dynamics of
explosive coronal events, geomagnetic substorms, solar wind coupling
to the magnetosphere, and magnetically confined fusion devices.  Early
models (\cite{Sweet58,Parker57,Petschek64}) and the predominance of
numerical work on magnetic reconnection ({\it e.g.}, \cite{Birn01})
have treated reconnection as two-dimensional.  However, naturally
occurring magnetic reconnection often begins in a localized region and
spreads in the direction perpendicular to the plane of reconnection.
For example, satellite observations of substorms in the magnetotail
identified a dawn-dusk asymmetry caused by localized reconnection
spreading in the westward direction (\cite{McPherron73,Nagai82}).  A
similar asymmetry was observed in the formation of arcades in the
solar corona (\cite{Isobe02}).  Capturing effects such as these
requires a fully three-dimensional treatment.

A number of numerical studies have addressed X-line spreading in the
direction of the current during quasi-two-dimensional reconnection.
Using a magnetic perturbation localized in the out-of-plane direction
in Hall-magnetohydrodynamics (Hall-MHD), it was found that the
localized reconnection signal propagates as a wave structure carried
by the electron current (\cite{Huba02,Huba03}).  By seeding
reconnection with large random magnetic perturbations in Hall-MHD
simulations, it was observed that reconnection develops into spatially
isolated structures that lengthen in the direction of the electron
current and that these small structures merge into larger scale
structures (\cite{Shay03}).  It was also suggested in this study that
spreading occurs in the direction of whichever species carries the
current, which need not be exclusively electrons.  Spreading by the
ions when they carry the current was observed in hybrid simulations
with localized resistivity (\cite{Karimabadi04}).  The result of these
works is that the reconnection X-line spreads in the out-of-plane
direction by the current carriers in the direction of the current
carriers (\cite{Lapenta06}).  \cite{Nakamura12} presented the first
systematic study to vary the fraction of current carried by each of
the species; the results confirmed that X-line spreading occurs due to
the current carriers.  The results are not dependent on the Harris
sheet geometry; \cite{Lukin11} observed X-line spreading in
simulations of island coalescence.  Note, each of these studies
primarily favored magnetotail applications, so they either treated
anti-parallel reconnection or reconnection with a weak out-of-plane
(guide) magnetic field compared to the background field.  X-line
spreading in a system without a guide field was recently observed in
laboratory experiments at the Magnetic Reconnection eXperiment (MRX),
and a physical mechanism for spreading by current carriers was
proposed (\cite{Dorfman12}).

Interestingly, experimental and satellite observations of systems with
a strong guide field reveal strikingly different behavior of X-line
spreading.  For example, experiments performed at the Versatile
Toroidal Facility (VTF) (\cite{Katz10,Egedal11}) exhibit reconnection
beginning in a localized region and spreading bi-directionally in the
out-of-plane (toroidal) direction at a speed consistent with the
Alfv\'en speed based on the guide field.  Another example is
bi-directional spreading (or elongating) of ribbons observed during
two-ribbon solar flares (\cite{Qiu09}), including the Bastille Day
flare (\cite{Qiu10}).  This presumably is related to spreading of the
looptop reconnection site where a sizable guide field is likely to be
present.  This spreading was also inferred to take place at the local
Alfv\'en speed.  Prominence eruptions in the corona have also been
observed to spread bi-directionally; this behavior was attributed to
magnetic reconnection propagating along the magnetic polarity
inversion line (PIL) (\cite{Tripathi06}).  In magnetospheric contexts,
observations of extended X-lines several Earth radii long at the
magnetopause (\cite{Phan00,Fuselier02}) and hundreds of Earth radii in
the solar wind (\cite{Phan06}) suggest that X-line spreading occurs in
these areas as well, although direct evidence of spreading is
prohibitively difficult with single- or even multi-point satellite
observations.  X-line spreading was also seen in three-dimensional
two-fluid simulations with a guide field (\cite{Schreier11}).

The existing observational data provide a clear indication that the
mechanism controlling X-line spreading strongly depends on the
strength of the guide field.  In the weak guide field limit, the
signal is transmitted by the current carriers; in the strong guide
field limit, the reconnection signal is transmitted by the magnetic
field as an Alfv\'en wave.  We hypothesize that, in general, the
X-line spreads in both directions at the speed of whichever mechanism
is faster for that direction.  In this paper, we present an estimate
of the critical guide field where the spreading mechanism changes and
confirm the theory with three-dimensional two-fluid numerical
simulations.

The layout of this paper is as follows.  A prediction of the critical
guide field at which the spreading mechanism changes from current
carriers to Alfv\'en waves is developed in Sec.~\ref{sec-theory}.  The
simulation setup and results are discussed in
Secs.~\ref{sec-simulations} and \ref{sec-results}, respectively.  A
discussion of the results and potential applications is in
Sec.~\ref{sec-discussion}.  We emphasize that we are considering
current sheets that are already thin, with large amounts of free
magnetic energy present.  The important topics of how the sheets
become thin and how the magnetic energy is stored is outside the scope
of this paper.

\section{Theory}
\label{sec-theory}

Here, we develop a prediction of the speed at which the X-line spreads
in each out-of-plane direction as a function of guide field and derive
the critical guide field at which the mechanism causing the spreading
changes from current carriers to Alfv\'en waves.  To do so, we make
the following simplifying assumptions.  We treat a
quasi-two-dimensional system, meaning that the equilibrium parameters
do not depend strongly on the direction normal to the reconnection
plane for all time.  We assume the current layer is flat, so that the
current sheet is either not curved or that the curvature does not
strongly contribute to the dynamics.  We assume the plasma parameters
are symmetric on either side of the current layer; asymmetries
(\cite{Cassak07d}) are not considered here.  Finally, we assume that a
single mode dominates the dynamics; in previous simulations, it was
shown that when multiple modes of reconnection occur, they can impede
the spreading of X-lines (\cite{Schreier11}).  This assumption is
valid at early times and in systems in which only a single mode is
present.

First, we estimate the spreading speed in each direction for each
spreading mechanism.  We begin with the speed due to the current
carriers.  From Amp\`ere's law, the current is ${\bf J} = c {\bf
  \nabla} \times {\bf B}/4 \pi$, where ${\bf B}$ is the magnetic
field.  For simplicity, we first assume the electrons carry the
out-of-plane current, so that the electron velocity is ${\bf v}_{e} =
-{\bf J}/ne$, where $n$ is the electron density and $e$ is the proton
charge.  Using a scaling argument, the electron speed $v_{eg}$ in the
out-of-plane direction is
\begin{equation}
  v_{eg} \sim \frac{cB_{rec}}{4 \pi n e \delta}, \label{veg}
\end{equation}
where $B_{rec}$ is the strength of the reconnecting magnetic field
upstream of the electron layer, $\delta$ is the thickness of the
current layer, and $g$ refers to the direction of the guide field.  As
has been previously established (\cite{Huba02, Shay03, Karimabadi04,
  Lapenta06, Lukin11, Nakamura12}), this is the X-line spreading speed
in the absence of a guide field.  In the strong guide field limit, the
observations suggest the spreading speed is the Alfv\'en speed
$c_{Ag}$ based on the guide field, given by
\begin{equation}
  c_{Ag} = \frac{B_{g}}{\sqrt{4 \pi m_{i} n}}, \label{cag}
\end{equation}
where $B_{g}$ is the strength of the guide field and $m_{i}$ is the
proton mass.

Our hypothesis is that the X-line spreading speed in the direction of the
electron out-of-plane flow, which we call $v_{Xe}$, is the larger of
$v_{eg}$ and $c_{Ag}$:
\begin{equation}
v_{Xe} = \max\{v_{eg},c_{Ag}\}. \label{vxe}
\end{equation}
From this, one can find the critical guide field $B_{crit,e}$ at which
the spreading mechanism changes, where the $e$ subscript denotes the
critical field for motion in the direction of the out-of-plane
electron flow.  Setting Eq.~(\ref{veg}) equal to Eq.~(\ref{cag}) and
solving for $B_{g}$ gives
\begin{equation}
  B_{crit,e} \sim B_{rec} \frac{d_{i}}{\delta}, \label{guidefield}
\end{equation}
where $d_{i} = c / \omega_{pi}$ is the ion inertial length and
$\omega_{pi} = (4 \pi n e^{2}/m_{i})^{1/2}$ is the ion plasma
frequency.  Since $\delta$ is typically less than $d_{i}$ as the
current is set by electron scales, we expect $B_{crit,e} > B_{rec}$,
although $B_{crit,e}$ is on the order of and slightly larger than the
reconnecting magnetic field strength upstream of the ion dissipation
region.

We perform a similar analysis for the spreading speed in the direction
of the ions $v_{Xi}$.  Since electrons carry the current, the ion
speed $v_{ig}$ in the out-of-plane direction is
\begin{equation}
  v_{ig} = 0. \label{ionflow}
\end{equation}  
Therefore, the X-line spreading speed in the direction of the ion
out-of-plane flow $v_{Xi} = \max\{v_{ig},c_{Ag}\}$ is given by the
Alfv\'en speed based on the guide field,
\begin{equation}
v_{Xi} = c_{Ag}. \label{vxi}
\end{equation}
Since $v_{ig} = 0$, the critical guide field $B_{crit,i}$ for
spreading in the direction of the ion out-of-plane flow is
\begin{equation}
  B_{crit,i} = 0. \label{guidefieldi}
\end{equation}

These results can be generalized to systems with both electrons and
ions carrying some of the current.  Following \cite{Nakamura12}, we
define the fraction of the total current $J_{z}$ carried by the ions
as $\alpha$, which is assumed known or measurable.  Letting $J_{iz} =
\alpha J_{z}$, one has $J_{ez} = (1 - \alpha) J_{z}$ so that $J_{z} =
J_{iz} + J_{ez}$.  By performing a similar analysis as before, one
finds the out-of-plane electron and ion flow speeds due to current
carrying are
\begin{eqnarray}
  v_{eg} & \sim & (1 - \alpha) \frac{cB_{rec}}{4 \pi n e \delta}, \nonumber \\
  v_{ig} & \sim & \alpha \frac{cB_{rec}}{4 \pi n e \delta},
\end{eqnarray}
which generalizes Eqs.~(\ref{veg}) and (\ref{ionflow}).  The X-line
spreading speeds in the direction of the electron and ion out-of-plane
flow are
\begin{eqnarray}
  v_{Xe} & = & \max\{v_{eg},c_{Ag}\}, \nonumber \\
  v_{Xi} & = & \max\{v_{ig},c_{Ag}\}, \label{spreadpred}
\end{eqnarray}
respectively, which generalizes Eqs.~(\ref{vxe}) and (\ref{vxi}).
Finally, the critical guide fields at which the mechanism for X-line
spreading changes from the current carriers to Alfv\'en waves in the
direction of electron and ion flows are given by
\begin{eqnarray}
  B_{crit,e} & \sim &  (1 - \alpha) B_{rec} \frac{d_{i}}{\delta}, \nonumber \\
  B_{crit,i} & \sim & \alpha B_{rec} \frac{d_{i}}{\delta}, \label{genbcrit}
\end{eqnarray}
respectively, which generalizes Eqs.~(\ref{guidefield}) and
(\ref{guidefieldi}).

\begin{figure}
\noindent\includegraphics[width=3.4in]{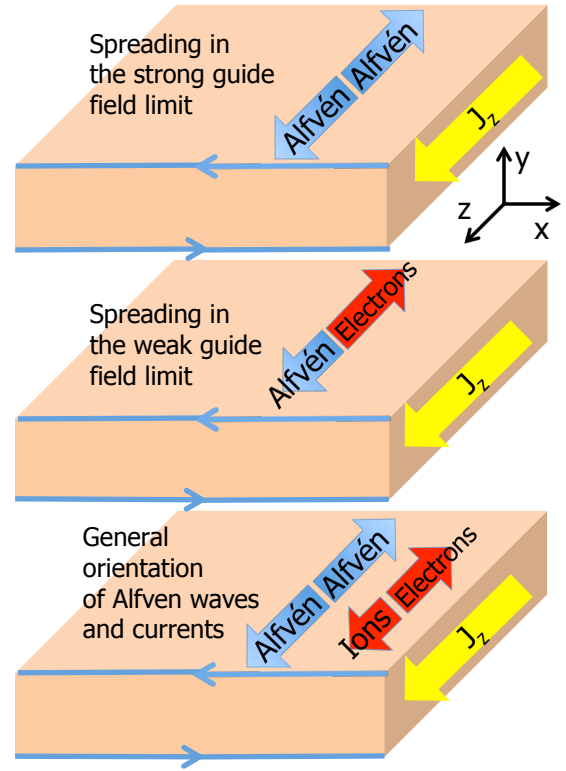}
\caption{\label{schematic} Schematic diagram showing the mechanisms
  that cause X-line spreading.  The thin blue arrows are the
  reconnecting magnetic field components, the yellow arrow is the
  total current.  The red arrows denote the speed of the current
  carriers; the thick blue arrows denote the speed of Alfv\'en waves
  along the guide field.  The top, middle, and bottom diagrams show
  the spreading mechanisms for strong, weak, and arbitrary guide field
  strengths.  In each case, X-line spreading occurs at the faster
  speed in each direction.}
\end{figure}

The predictions derived here are summarized pictorially in
Fig.~\ref{schematic}, where the current is depicted by the yellow
arrows and the reconnecting magnetic fields are the thin blue lines.
The thick arrows denote the speeds of the current carriers (in red)
and the Alfv\'en speed (in blue) in each out-of-plane direction.  The
top, middle, and bottom plots show the results for strong, weak, and
arbitrary guide field strengths, respectively.  In each case, the
X-line spreading speed is the longer of the arrows on either side.  We
point out that there is nothing preventing the spreading mechanisms
from being different in the two directions, {\it i.e.}, Alfv\'en waves
in one direction and current carriers in the other, if that is what
Eq.~(\ref{genbcrit}) dictates for the system parameters.

\section{Simulation Setup}
\label{sec-simulations}

To test the predictions on X-line spreading, three-dimensional
numerical simulations are performed using the two-fluid code
F3D~(\cite{Shay04}).  The code updates the continuity, momentum, and
induction equations with the generalized Ohm's law including electron
inertia.  Magnetic fields and densities are normalized to arbitrary
values $B_{0}$ and $n_{0}$.  Velocities are normalized to the Alfv\'en
speed $c_{A0}= B_{0}/(4 \pi m_{i}n_{0})^{1/2}$.  Lengths are
normalized to the ion inertial length $d_{i0}=c/\omega_{pi0} = (m_{i}
c^{2} / 4 \pi n_{0} e^{2})^{1/2}$.  Times are normalized to the ion
cyclotron time $\Omega_{ci0}^{-1} = (e B_{0} / m_{i} c)^{-1}$,
electric fields to $E_{0} = c_{A0}B_{0}/c$, and temperatures to $T_{0}
= m_{i} c_{A0}^{2}$.

Simulations are performed in a three-dimensional domain of size $L_{x}
\times L_{y} \times L_{z} = 51.2 \times 25.6 \times 256.0 \textrm{ }
d_{i0}$, where $x$ is the direction of the oppositely directed field,
$y$ corresponds to the inflow direction if the simulations were
two-dimensional, and $z$ is the direction of the initial current.  The
plasma is assumed to be isothermal and there is no resistivity ($\eta
= 0$).  Boundaries in all three directions are periodic, but the
system is long enough in the $z$ direction that the periodic
boundaries do not affect the dynamics on the time scales of import to
the present study.

For simplicity, the simulations have the electrons carrying all of the
initial current ({\it i.e.,} $\alpha = 0$).  The electron inertia is
$m_{e}= m_{i}/25$.  In previous simulations with this electron mass
(and confirmed in the simulations here), it has been observed that the
current layer thickness $\delta$ thins down to the electron inertial
scale $d_{e} = 0.2 \ d_{i}$ and the reconnecting magnetic field at the
electron layer is $B_{rec} \simeq 0.4 \ B_{0}$ (\cite{Jemella03}).
Substituting this into Eq.~(\ref{guidefield}), we predict a critical
guide field of
\begin{equation}
  B_{crit} \simeq 2 B_{0}. \label{bcritsims}
\end{equation}
Therefore, we can test the theory by running a series of simulations
in which the initial guide fields are $B_{g} = 0,0.5,1,1.5,2,2.5$ and
3.  Note, the scaling $\delta \sim d_{e}$ and $B_{rec} \sim 0.4 \
B_{0}$ may or may not be representative of naturally occurring
reconnection; care should be taken to investigate this for particular
applications.

The initial configuration is a double tearing mode with two Harris
sheets, $B_{x0}(y) = \tanh[(y + L_{y}/4)/w_{0}] - \tanh[(y -
L_{y}/4)/w_{0}] - 1$, with uniform initial temperature $T = 1$ and a
non-uniform plasma density to balance total pressure.  Here, $w_{0} =
0.4 \ d_{i0}$ is the initial current layer thickness.  We choose this
thickness to be comparable to the smallest value of the ion Larmor
radius $\rho_{s} = c_{s} / \Omega_{ci} = \sqrt{T} / B_{g} \simeq
0.33$, where $c_{s}$ is the sound speed and the latter expression is
written in normalized units.  This scale is the Hall scale in the
presence of a strong guide field (\cite{Zakharov93,Rogers01}).  It is
worth noting that the Hall scale increases smoothly from $\rho_{s}$ to
$d_{i}$ as the guide field is decreased to zero, which follows from a
linear analysis of Hall-MHD waves (\cite{Rogers01}).  Consequently,
the smaller guide field simulations start with a current sheet that is
thin relative to the Hall scale, and should onset rapidly.  As the
guide field is increased, the time to onset should increase and it is
expected that a hyper-resistive phase of reconnection will occur
before onset.  This behavior will not adversely impact our study, as
we will separate out the times for which Hall reconnection is
dominant.

We employ a grid scale of $\Delta x \times \Delta y \times \Delta z =
0.05 \times 0.05 \times 1.0\textrm{ } d_{i0}$.  Using a stretched grid
in the out-of-plane direction has been done before (\cite{Shay03}),
and is acceptable since the in-plane kinetic-scale dynamics is on
smaller scales than the out-of-plane dynamics.  To ensure the
stretched grid scale in the out-of-plane direction does not play a
role in the numerics, some simulations are confirmed by comparison
with simulations with $\Delta z = 0.5 \ d_{i0}$.  All equations employ
a fourth-order diffusion with coefficient $D_{4x} = D_{4y} = 2.5\times
10^{-5}$ in the $x$ and $y$ directions.  In the out-of-plane direction
the fourth-order diffusion coefficient $D_{4g}$ depends on the speeds
in the out-of-plane direction.  For $B_{g} \le 2.0$ the fourth-order
diffusion coefficient is $D_{4g} = 0.064$ and for $B_{g} = 2.5$ and
$3.0$ the fourth-order diffusion coefficient is $D_{4g} = 0.081$ and
$0.097$, respectively.  The values of $D_{4g}$ were tested by varying
the value by a factor of two to ensure that $D_{4g}$ does not play a
significant role in the dynamics.

The inclusion of a guide field in these simulations changes the nature
of reconnection relative to previous work on X-line spreading.  In
three-dimensional periodic domains, it is well established that the
linear tearing instability is excited where ${\bf k} \cdot {\bf B}_{0}
= 0$, where ${\bf B}_{0} = B_{x0} {\bf \hat{x}} + B_{z0} {\bf
  \hat{z}}$ is the equilibrium magnetic field and ${\bf k} = k_{x}
{\bf \hat{x}} + k_{z} {\bf \hat{z}}$ is the wave vector of the mode.
The periodic domain enforces that $k_{x} = 2\pi m/L_{x}$ and $k_{z} =
2 \pi n/L_{z}$, where $m$ and $n$ are integer mode numbers which
specify the number of X-lines in the $x$ and $z$ direction,
respectively.  In the absence of a guide field, this condition is only
satisfied where $B_{x0} = 0$.  With a guide field, it is satisfied
wherever $q(y) = L_{x} B_{z}(y) / L_{z} B_{x}(y) = m / n$ is a
rational number, where $q(y)$ is the safety factor well known in
fusion applications.  The $y$ locations where ${\bf k} \cdot {\bf
  B}_{0} = 0$ is satisfied are called rational surfaces, and for the
equilibrium profile here, the modes are displaced from where $B_{x0} =
0$ by a distance $y_{s} = w_{0} \tanh^{-1}(n L_{x} B_{g} / m L_{z}
B_{0})$.  Thus, modes in our simulations are excited on multiple
rational surfaces.

\begin{figure}
\noindent{}\includegraphics[width=3.4in]{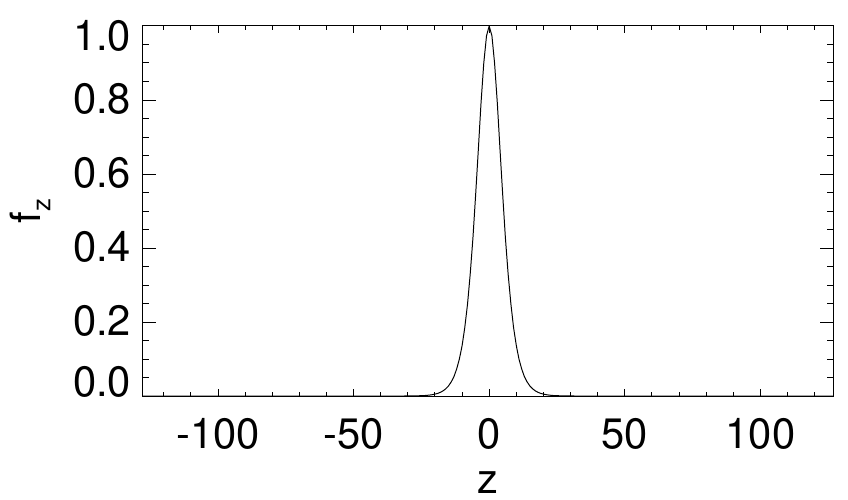}
\caption{\label{fzz} The envelope $f_{z}(z)$ used to localize the
  magnetic perturbation in the out-of-plane direction $z$.}
\end{figure}

Reconnection is seeded using a coherent magnetic perturbation
localized in the out-of-plane direction of the form
\begin{equation}
  B_{y1} = \sum \limits_{k_{x},k_{z}}\widetilde{B}_{1}
  \sin(k_{x}x + k_{z}z)f_{z}(z),  \label{perturbation}
\end{equation}
where $\widetilde{B}_{1}=0.1$.  Here, $f_{z}(z)$ is an envelope that
localizes the perturbation in the out-of-plane direction and is given
by $f_{z}(z) = \{\tanh [(z+w_{0z})/6] - \tanh[(z - w_{0z})/6]\} / 2$.
We use $w_{0z} = 1$; a plot of $f_{z}(z)$ is in Fig.~\ref{fzz}.
Random magnetic perturbations that range from $m,n = 0 \textrm{ to
}20$ with small amplitude $0.02 \ B_{0}$ are included with the initial
conditions to break symmetry so that secondary islands are ejected.

In early simulations, we initially excite only the $(m,n)=(1,0)$ mode
in Eq.~(\ref{perturbation}).  Even though this mode is the strongest
perturbation, oblique modes with $n \neq 0$ grow from the noise and
dominate the reconnection.  This is consistent with recent
particle-in-cell (PIC) simulations~(\cite{Daughton11}) and linear
theory~(\cite{Baalrud12}).  Oblique modes in reconnection have been
observed many times in fusion applications (see {\it e.g.},
\cite{Grasso07}).  In light of these results, we include oblique modes
in Eq.~(\ref{perturbation}) and compare the results with the original
simulations.  The values of $m$ and $n$ are chosen so that the
displacement $y_{s}$ is less than $w_{0}$.  In this study, $m=1$ for
all simulations and $n$ ranges from $0$ to $3$.  Initially exciting
oblique modes has no noticeable effect on the results on the
development of reconnection.  Thus, the results of this study are
expected to be independent of the modes used to seed reconnection.
Note, although the modes are oblique, they are still quasi
two-dimensional until they start interacting strongly.  It was shown
(\cite{Schreier11}) that interacting oblique modes can prevent X-lines
from spreading, so we focus on times early enough in the evolution
that oblique mode interactions have not yet occurred.

\begin{figure*}
\indent{}\includegraphics[width=6.8in]{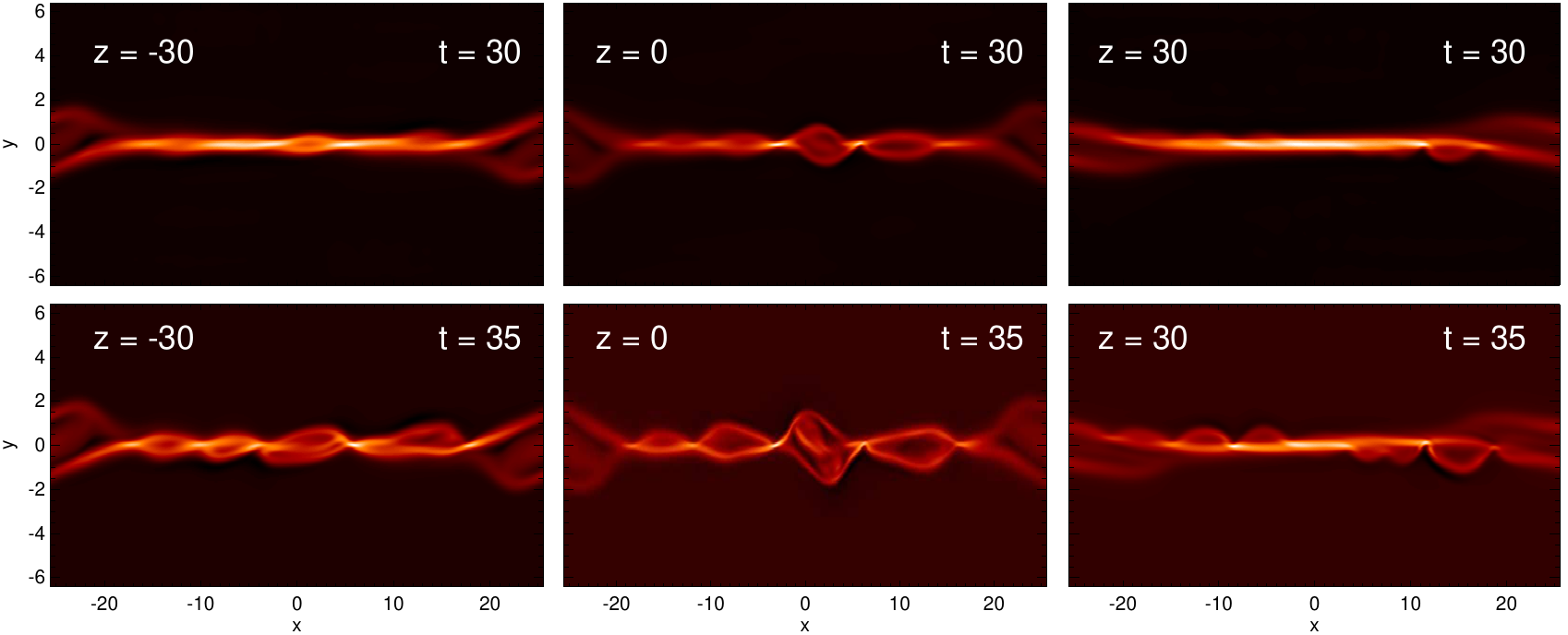}
  \caption{\label{curzxy} Cuts at different values of $z$ of the
    out-of-plane current $J_{z}$ for the $B_{g} = 3$ simulation.  At
    $t=30$ (top row), Hall reconnection is developing at $z=0$ but not
    at $z=\pm 30$.  At $t=35$ (bottom row), Hall reconnection has
    developed fully at $z=0$ and is developing at $z=\pm 30$. }
\end{figure*}

\section{Results}
\label{sec-results}

To ensure numerical feasibility of the simulations in three
dimensions, we benchmark the simulations in two dimensions.  The
simulations are evolved from $t = 0$ until nonlinear reconnection
develops.  In two dimensions, symmetry dictates that the $n = 0$ mode
is the only excited mode. The reconnection rate $E$, measured as the
time rate of change of the difference in magnetic flux between the
X-line and O-line during a quasi-steady period, is approximately
0.08-0.1 for all simulations.  Also, as expected, the time until Hall
reconnection begins increases as the guide field increases since
$w_{0}$ is held fixed, and there is a brief hyper-resistive
reconnection phase before onset for stronger guide fields.

In the three-dimensional simulations, the evolution at $z = 0$ is very
similar to what is observed in two dimensions: the time scale of the
development of reconnection is comparable, and a hyper-resistive phase
precedes Hall reconnection.  The reconnection rates can be compared,
as well.  The reconnection rate in three-dimensions is measured by
taking a cut of the $z$ component of ${\bf v} \times {\bf B}$ in the
$y$-direction across the X-line; far from the current sheet, it
asymptotes to the reconnection electric field in a steady-state.  The
reconnection rates are in the 0.08-0.1 range, comparable to the
two-dimensional results.  One noticeable difference, as discussed
earlier, is that oblique modes dominate over $n = 0$ modes in
three-dimensions.

Each of the three-dimensional simulations display some form of X-line
spreading.  This can be seen qualitatively in Fig.~\ref{curzxy} for
the simulation with $B_{g} = 3$.  The out-of-plane current $J_{z}$ at
time $t=30$ (top row) and at time $t = 35$ (bottom row) is displayed
at three different out-of-plane positions: $z = -30, 0,$ and $30$ from
left to right.  At the earlier time $t = 30$, a transition to fast
(Hall) reconnection at multiple sites at $z = 0$ has occurred,
consistent with the development of multiple oblique modes.  At $z =
\pm 30$, the reconnection is still hyper-resistive.  At the later time
$t = 35$, the current sheet at all three positions in $z$ has
developed multiple Hall reconnection X-lines.  Thus, the Hall
reconnection signal propagates bi-directionally from $z = 0$ for
$B_{g} = 3$.  It is worth noting that the multiple oblique mode
reconnection seen here is consistent with previous simulations, and
the reason multiple X-lines appear despite the $m=1$ mode being the
dominant mode is that there are multiple modes simultaneously excited
on different rational surfaces.

\begin{figure*}
\noindent{}\includegraphics[width=6.8in]{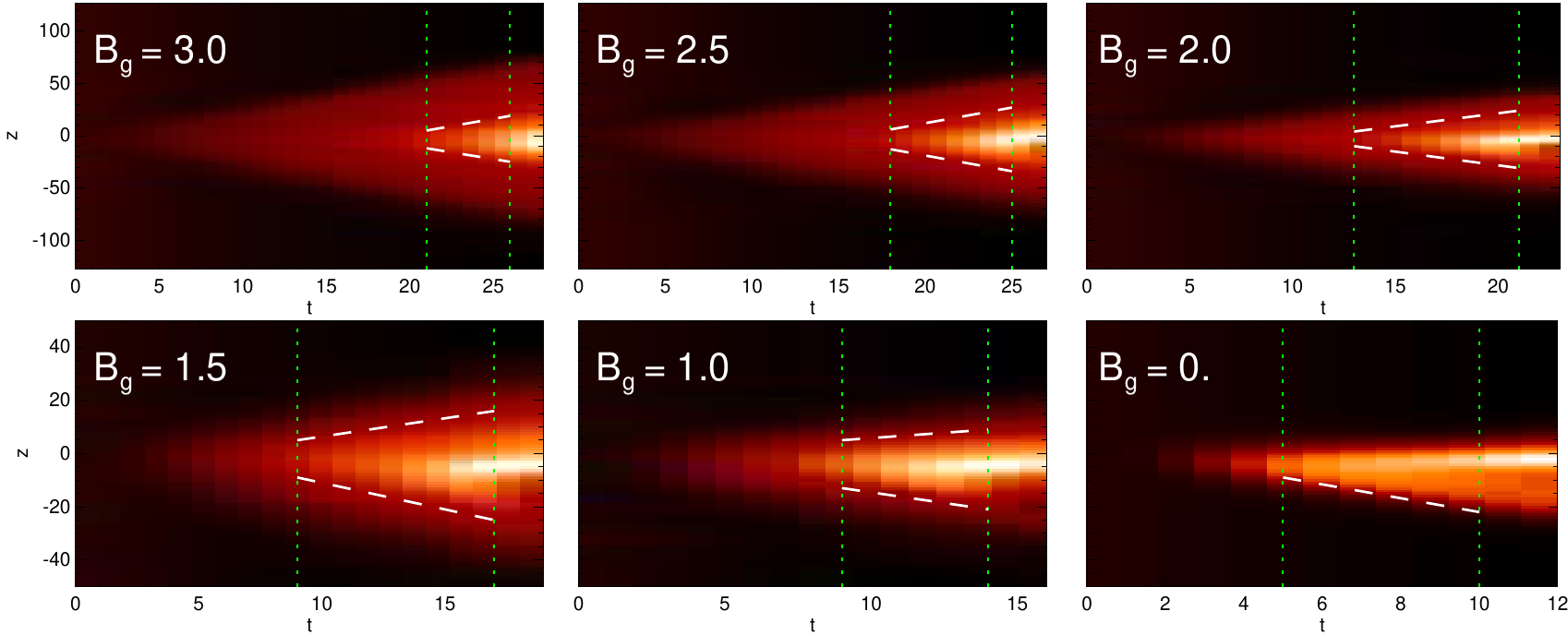}
  \caption{\label{curzarray} Stack plots of $J_{max}(z)$ as a function
    of $t$ and $z$.  The vertical dashed green lines indicate the
    range of time over which the spreading speed is measured, $t_{i}$
    and $t_{f}$.  The dashed white lines indicate the extent in $z$ of
    the X-line; their slope gives the speed of the spreading.  Note,
    the images in the bottom row are on a different scale in $z$ than
    those in the top row.}
\end{figure*}

To quantify the speed at which the X-line spreads, we must develop a
systematic way to determine the extent of the reconnection region.  As
Hall reconnection develops, the out-of-plane current $J_{z}$ at the
X-line becomes noticeably higher than regions where reconnection is
hyper-resistive.  For each slice in $z$, we measure the maximum
out-of-plane current, which we call $J_{max}(z)$.  These maximum
values of the current correspond to the location of the X-line for
each position in $z$.  The extent of the X-line can then be readily
seen in a stack plot of $J_{max}(z)$ as a function of $t$.  Stack
plots for all six initial guide fields in this study are displayed in
Fig.~\ref{curzarray}.  As mentioned earlier, the plots only cover
early times when the three-dimensional X-line structure is well
defined because the interaction of oblique modes make defining the
X-line structure prohibitive.

The bright white regions in Fig.~\ref{curzarray} correspond to the
strongest currents and, thus, the Hall reconnection X-lines.  The
dimmer areas outside of the white dashed lines (the red to black
colors) indicate the region undergoing hyper-resistive reconnection.
As expected, the $B_{g} = 0$ simulation onsets almost immediately
without a hyper-resistive phase since $w_{0} < d_{i}$, while the onset
time increases as the guide field increases, leading to a longer
hyper-resistive phase.  Both phases of reconnection spread in the $z$
direction as time evolves; we focus on the Hall reconnection X-lines
in the present study.

From Fig.~\ref{curzarray}, the qualitative differences in the nature
of the spreading as a function of the guide field can readily be seen.
For the strong guide field simulations $B_{g} \ge 2$, the X-line
spreads symmetrically about $z=0$ (top row), which is consistent with
our expectations for the strong guide field regime from
Eq.~(\ref{bcritsims}).  However, for simulations with a guide field
weaker than the predicted condition $B_{g}<2$, we observe different
spreading behavior in the $+z$ and $-z$ directions (bottom row).  For
the $B_{g} = 1.5$ case, there is bi-directional spreading, as observed
in the stronger guide field runs, but the spreading is not symmetric
about $z=0$.  The spreading in the $-z$ direction appears marginally
faster than in the $+z$ direction.  These differences are further
amplified in the $B_{g} = 1$ simulation.  With no guide field ($B_{g}
= 0$), spreading occurs primarily in the $-z$ direction, with
negligible spreading in the $+z$ direction.  Since $J_{z}$ is in the
$+z$ direction for this reconnection site, the propagation is in the
direction of the electron out-of-plane flow, consistent with previous
work (\cite{Huba02}).

To make this quantitative, we measure the spreading speed of the
X-line after Hall reconnection begins by finding the length of the
X-line in the out-of-plane direction; its time rate of change between
an initial and final time is the spreading speed.  To do so, we note
that the reconnection rate during hyper-resistive reconnection never
exceeds 0.01.  We observe that when the out-of-plane Hall electric
field $E_{Hg} = ({\bf J} \times {\bf B})_{g} / nec$ in a cut in the
$y$ direction through the X-line exceeds 0.01, the reconnection has
begun its transition to Hall reconnection.  We also note empirically
that $J_{max}$ at the time of this transition is always close to 3.3,
which is robust for all the simulations performed here.  Thus, we take
Hall reconnection as occurring when $J_{max}$ exceeds a threshold
value of $J_{thresh} = 3.3$.

The time frame over which the spreading speed is measured is defined
as follows.  The initial time $t_{i}$ is defined as the earliest time
that $J_{max}$ exceeds $J_{thresh}$ over the entire range from $z =
\pm 5$.  This range of $z$ is chosen because the initial magnetic
perturbation that seeds the X-lines is localized in this region, so
genuine spreading not being influenced by the growth of reconnection
inside the initially perturbed region requires the signal to leave
this range in $z$.  The final time $t_{f}$ is defined for each
simulation as the latest time in the evolution before multiple oblique
modes interact; this assessment is done visually by finding where the
current develops complicated structure as seen in Fig.~\ref{curzxy}.
The length of the X-line at a given time is defined as the extent in
$z$ for which $J_{max}$ exceeds $J_{thresh}$.  The spreading speed is
calculated as the difference of the length of the X-line between
$t_{f}$ and $t_{i}$ divided by the time difference.

\begin{figure}
\noindent{}\includegraphics[width=3.4in]{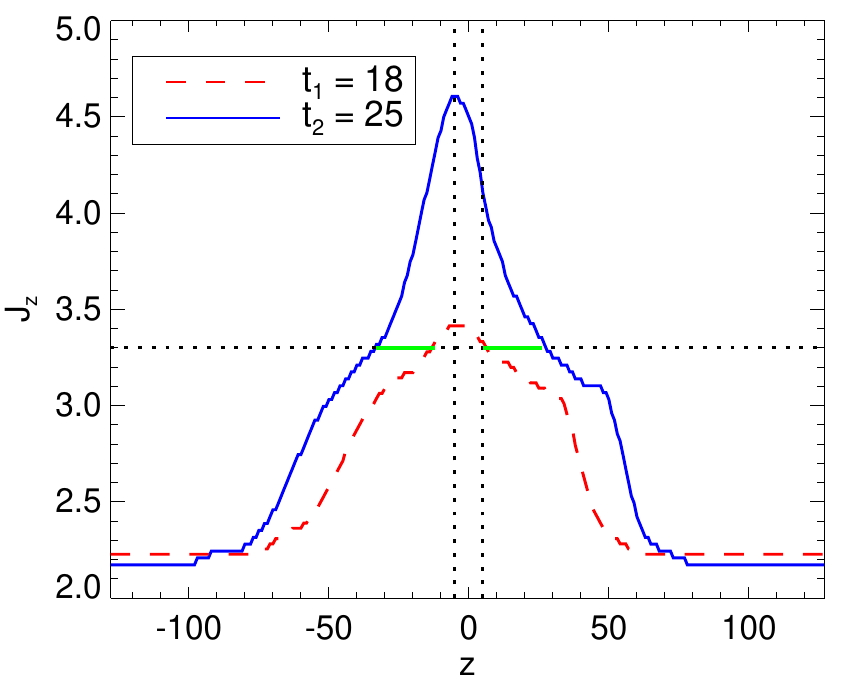}
\caption{\label{curzcuts} Cuts of the stack plot for the $B_{g} = 2.5$
  case plotted in Fig.~\ref{curzarray}.  The dashed (red) line and
  solid (blue) line are at $t_{i}=18$ and $t_{f}=25$, respectively.
  The horizontal dotted line is at $J_{thresh}$ as defined in the
  text.  The vertical dotted lines mark $z = \pm 5$, the approximate
  extent of the initial magnetic perturbation.  The green line denotes
  the change in length of the X-line between $t_{i}$ and $t_{f}$.}
\end{figure}

An example of this procedure is presented in Fig.~\ref{curzcuts},
where representative data for the $B_{g} = 2.5$ simulation is shown.
The initial time is $t_{i} = 18$, which is the earliest time that
$J_{max} > J_{thresh}$ everywhere between $z = \pm 5$, as shown by the
red dashed line.  The final time is taken to be $t_{f} = 25$; a plot
of $J_{max}(z)$ at $t_{f}$ is shown as the blue line.  The horizontal
dotted line marks the current threshold $J_{thresh} = 3.3$ and the
vertical dotted lines mark the boundary of $z=\pm 5$.  The change in
length between the two times is the distance between the curves at
$J_{thresh}$, marked by the green line segments.  The lengths and
speeds are calculated separately for the $\pm z$ direction because the
speeds in the two directions may be different depending on the
strength of the guide field.  For the $B_{g} = 2.5$ simulation, the
change in length in the $+z$ and $-z$ directions are 21 and 20
$d_{i0}$, respectively, and dividing by the time difference gives
speeds of $v_{Xi} = 3.0 \ c_{A0}$ for the speed in the $+z$ direction
(the direction of ion out-of-plane flow) and $v_{Xe} = 2.9 \ c_{A0}$
for the speed in the $-z$ direction (the direction of electron
out-of-plane flow).

The initial and final times $t_{i}$ and $t_{f}$ for each simulation
are illustrated in Fig.~\ref{curzarray} as the vertical dotted green
lines.  The dashed white lines connect the extent of the X-line at the
initial and final times.  By inspection, one can see that the
technique we employ to measure the extent of the X-line appropriately
captures the evolution of the X-line length.  Also, since the region
of stronger current is rather straight between the beginning and final
times, this implies the spreading speed is approximately constant in
time.

\begin{figure}
\noindent{}\includegraphics[width=3.4in]{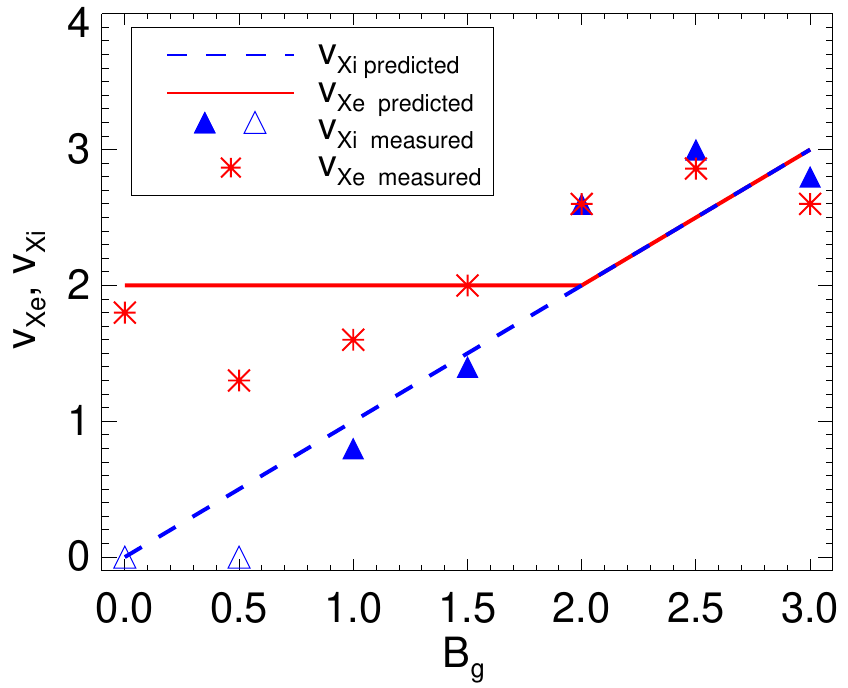}
\caption{\label{spread} Spreading speeds $v_{Xi}$ and $v_{Xe}$ as a
  function of guide field $B_{g}$.  The red asterisks are the measured
  values of $v_{Xe}$; the solid red line is the prediction from
  Eq.~(\ref{vxe}).  The solid blue triangles are the measured values
  of $v_{Xi}$; the open triangles are for simulations for which no
  spreading was measured.  The dashed blue line is the prediction from
  Eq.~(\ref{vxi}).}
\end{figure}

The measured X-line spreading speeds $v_{Xe}$ and $v_{Xi}$ are
calculated as the time rate of change of the length of the X-line,
which is equivalent to the slope of the white dashed lines in
Fig.~\ref{curzarray}.  The results for the spreading speed in both
directions are plotted as a function of guide field $B_{g}$ in
Fig.~\ref{spread}.  The measured value of the spreading speed is given
by the solid blue triangles for $v_{Xi}$ and the red stars for
$v_{Xe}$.  Note, $v_{Xi}$ for $B_{g} = 0.0$ and 0.5 is plotted as zero
as the hollow blue triangles.  This is because the Hall reconnection
signal is found to not extend past $z = \pm 5$ for either simulation
during the time considered.

To compare these results to the theory, note that the electrons carry
all of the out-of-plane current in the $-z$ direction in our
simulations.  Therefore, in the weak guide field regime $B_{g} < 2$,
Eq.~(\ref{vxe}) predicts that the spreading speed in the direction of
the electron current $v_{Xe}$ is the speed of the electrons given in
Eq.~(\ref{veg}), which is independent of $B_{g}$.  When $B_{g} \ge 2$,
the spreading speed is determined by the Alfv\'en speed given by
Eq.~(\ref{cag}), which increases linearly with $B_{g}$.  The predicted
speed of X-line spreading in the direction of the ion current $v_{Xi}$
is the Alfv\'en speed due to the guide field, as given by
Eq.~(\ref{vxi}), which increases linearly with $B_{g}$ for all guide
field strengths.

The predicted spreading speeds $v_{Xe}$ and $v_{Xi}$ are depicted in
Fig.~\ref{spread} by the solid red line and the dashed blue line,
respectively.  Qualitatively, the data reveal that the nature of
X-line spreading is sensitive to the strength of the guide field.  To
interpret this more quantitatively, we first discuss the estimated
uncertainties in our speed measurements.  If we use a higher value of
the current threshold $J_{thresh}$, the spreading speed changes on the
order of 15-20\%, which we take as the uncertainty.  We note that for
the large guide field runs $B_{g} \ge 2$, the speeds in either
direction are within the uncertainties of each other.  However, for
$B_{g} < 2$, the speeds in either direction are separated by more than
their uncertainty.  These two results suggest that the spreading
mechanism is the same in both directions for $B_{g} \ge 2$ and is
different in either direction for $B_{g} < 2$, which quantitatively
agrees with Eq.~(\ref{bcritsims}).

For the absolute spreading speeds, when the estimated uncertainties
are taken into account, the measured values agree pretty well with the
predicted speeds.  It is unexpected that the $B_{g} = 3$ speeds are
slower than $B_{g} = 2.5$, but both are within the uncertainties of
the predicted value.  Also, it is expected that for $B_{g} = 0.5$, a
non-zero value could be obtained if there had been a longer time
before the oblique modes started interacting.  Therefore, we conclude
that data in Fig.~\ref{spread} quantitatively support the theory
presented in Sec.~\ref{sec-theory}.

In conclusion, the mechanism of X-line spreading in the out-of-plane
direction is qualitatively different depending on the strength of the
guide magnetic field.  For $B_{g} \ge B_{crit,e}$, X-line spreading
occurs bi-directionally along the guide field at the Alfv\'en speed.
For $B_{g}\le B_{crit,e}$, X-line spreading occurs bi-directionally
along the guide field, but the spreading speed in the direction of the
current carriers is the speed of the current carriers and in the
direction opposite of the primary current carriers the spreading speed
is the Alfv\'en speed.  Measurements of X-line spreading for the
hyper-resistive reconnection that precedes Hall reconnection agree
with the results obtained from measuring the Hall reconnection
spreading (not shown).  Therefore, the main result of this study
applies both to Hall and hyper-resistive reconnection in a two-fluid
model.

\section{Discussion}
\label{sec-discussion}

In summary, the mechanism of X-line spreading in the out-of-plane
direction is qualitatively different for strong guide magnetic fields
than it is for weak guide fields.  For weak guide fields, the
reconnection signal is propagated by the current carriers, as has
previously been established; for strong guide fields, the reconnection
signal is propagated by Alfv\'en waves along the guide field.  In
general, the spreading speed in either out-of-plane direction is given
by the maximum of the speed of the current carriers in that direction
and the Alfv\'en speed based on the guide field, as given by
Eq.~(\ref{spreadpred}).

Because the changeover from one spreading mechanism to the other is
abrupt, there is a critical guide field strength (for each direction)
at which the nature of the spreading switches.  This critical field
depends only on the strength of the reconnecting magnetic field, the
ion inertial scale, the thickness of the electron dissipation region,
and the fraction of the current carried by each species, as given by
Eq.~(\ref{genbcrit}).  When the guide field $B_{g}$ exceeds the
critical field, the spreading is due to Alfv\'en waves; when it is
smaller, the spreading is due to the current carriers.  The weak guide
field result is consistent with previous numerical work of X-line
spreading~(\cite{Huba02,Shay03,Karimabadi04,Lapenta06,Nakamura12}),
but the new result generalizes the predictions to include a guide
field.

The present results may be relevant for interpreting observations of
reconnection in many settings.  For example, in laboratory
experiments, X-line spreading has been observed to be bi-directional
and at the Alfv\'en speed in the strong guide field limit
(\cite{Katz10}) and uni-directional in the small guide field limit
(\cite{Dorfman12}).  These results are consistent with the results of
the present study.

Another potential application is for solar flares.  Two-ribbon flare
evolution is marked by the ribbons moving apart from each other as
time evolves, which is interpreted as newly reconnected field lines
piling on top of previously reconnected field lines.  In addition to
this behavior, bi-directional spreading or elongation of the ribbon in
the direction parallel to the ribbons along the polarity inversion
line has been observed (\cite{Qiu09}).  It was shown that the
spreading speed was consistent with the Alfv\'en speed (\cite{Qiu09}).
Since the reconnection driving the flare is most likely to have a
sizable guide field, the present results suggest that this type of
bi-directional spreading at the Alfv\'en speed would be expected.

The Bastille Day flare exhibits this spreading, as well
(\cite{Qiu10}).  From geometrical considerations of the magnetic
fields of the flare loops, it was argued that the guide field was of
comparable size as the reconnecting field, with $B_g \simeq 0.4-1.2$
times the reconnecting field (\cite{Qiu10}).  We can check this using
the present results and the observed properties of the spreading.
From the observations, the spreading speed ranged between $30 - 70
\textrm{ km/s }$ (\cite{Qiu10}).  Let us assume the spreading is
governed by Alfv\'en waves.  Assuming an average density of $n =
10^{13} {\rm \ cm}^{-3}$ (\cite{Qiu10}), the guide field ranges from
$B_{g} \simeq 15-100$ G using Eq.~(\ref{cag}).  The motion of the
ribbons normal to the ribbons was $20 \textrm{ km/s}$ (\cite{Qiu10}),
which is expected to be correlated to the inflow speed at the
reconnection site.  Since the inflow speed is often taken to be 0.1 of
the Alfv\'en speed based on the reconnecting field, the reconnecting
field strength $B_{rec} \simeq 140$ G.  These results suggest the
guide field is about 0.1-0.7 of the reconnecting field.  Despite the
large uncertainties, the two techniques give similar results.  This
analysis is obviously oversimplified and merely presented as an
example of how the results can be used, but it is hoped that future
work will allow for a meaningful assessment of the relative strengths
of the guide and reconnecting fields.  The reason this may be useful,
as emphasized by \cite{Qiu10}, is that the strength of the guide field
is known to influence the production of secondary islands
(\cite{Drake06b}), and it has been suggested that the presence of
secondary islands (plasmoids) is important for particle acceleration
(\cite{Drake06c}).

Another interesting application is for reconnection in the solar wind,
where reconnection X-lines extending hundreds of Earth radii have been
reported (\cite{Phan06}).  One can ask whether the spreading of
reconnection in the out-of-plane direction could allow the X-line to
be that long.  To estimate the size of X-lines, assume that
reconnection begins close to the Sun with an initially small finite
length in the out-of-plane direction.  Suppose the reconnection site
convects out with the solar wind at a speed $v_{SW}$.  (For
simplicity, this calculation ignores variations in solar wind speed,
magnetic field strength, and plasma density as a function of distance
from the Sun.)  Then, the time it takes to get to a position $r_{f}$
away from the Sun is $t \sim r_{f}/v_{SW}$.  If the speed of the
spreading of the X-line is $v_{X}$, then the extent $L$ of the X-line
at $r_{f}$ is $L \sim v_{X} t \sim r_{f} v_{X} / v_{SW}$, which gives
the upper limit on the length of the X-line that could arise in the
solar wind.

One can test the implications of this from the observations of the
reconnection event in the \cite{Phan06} study, where a solar wind
speed is inferred to be $v_{SW} = 340 \textrm{ km/s}$.  The satellite
observations occurred near the Earth, so $r_{f} \simeq 1 {\rm \ AU}
\simeq 2.3 \times 10^{4} R_{E}$.  The Alfv\'en speed based on a guide
field of strength $B_{g} = 4$ nT and density $n = 20 \textrm{ cm}^3$
(\cite{Phan06}) is $c_{Ag} = 19 \textrm{ km/s}$.  If we take this as
the spreading speed $v_{X}$, then the maximum length of the X-line is
$L \sim r_{f} v_{X}/v_{SW} \sim 1.3 \times 10^{3} R_{E}$.  This
exceeds the length of the X-line reported by \cite{Phan06}, which was
390 $R_{E}$.  In this event, the strength of the guide field was 0.35
of the reconnecting field, so the Alfv\'en speed is the slower of the
two velocities and the extent of the X-line if spreading is due to
current carriers is even longer.  Thus, while this calculation assumes
that the reconnection proceeds at short distance from the Sun, it
gives an indication that it is not impossible to achieve reconnection
X-lines of the lengths reported by \cite{Phan06}.  Further work is
necessary to make more careful comparisons of the theory to data.

We conclude by collecting some assumptions of this work.  We treat our
system as quasi-two-dimensional, meaning any variation in the system
in the direction of the current is negligible.  The current sheet in
all simulations performed are initially thin, meaning that free
magnetic energy has already been stored.  The plasma parameters across
the current sheet are assumed symmetric.  The simulations employ a
two-fluid model, which does not fully capture electron scale physics.
This may make quantitative changes to our results (such as the
estimate of $\delta$ and the size of $B_{rec}$), but we do not expect
qualitative changes to the theoretical results.  Also, the simulations
are isothermal and contain no thermal conduction.

Another main assumption is that only a single mode is dominating the
dynamics.  However, the role of multiple oblique modes can play an
important part of the dynamics of the spreading process.  As seen in
our simulations, the X-line structure is identifiable at early times
but as the complicated nature of the oblique modes develop, the X-line
structure break up due to the interaction between the current sheet.
The interaction of oblique modes can impede X-line
spreading~(\cite{Schreier11}).  More work is necessary on this topic.

\begin{acknowledgments}
  The authors would like to thank J.~Egedal, J.~T.~Gosling, H.~Isobe,
  M.~G.~Linton, V.~S.~Lukin, T.-D.~Phan, J.~Qiu, M.~A.~Shay, and
  D.~G.~Sibeck for helpful conversations.  Support by NSF Grants
  PHY-0902479 and AGS-0953463 and NASA Grant No.~NNX10AN08A is
  gratefully acknowledged.  This research used resources of the
  National Energy Research Scientific Computing Center, which is
  supported by the Office of Science of the U.~S.~Department of Energy
  under Contract No.~DE-AC02-05CH11231.
\end{acknowledgments}

\end{article}

\end{document}